\documentclass[12pt,preprintnumbers,amsmath,amssymb,superscriptaddress,nofootinbib]{revtex4}

\usepackage{graphicx}
\usepackage{dcolumn}
\usepackage{bm}
\usepackage{graphics}
\usepackage{amssymb}
\usepackage{amscd}
\usepackage{afterpage}
\usepackage{float,times}
\usepackage{subfigure}
\usepackage{rotating}
\usepackage{multirow}
\usepackage{epsfig}
\usepackage{theorem}
\usepackage{moreverb}
\usepackage{euscript}
\usepackage{psfrag}
\begin{document}

{\hbox to\hsize{\hfill September 2007 }}
\title{A solution to the hierarchy problem from an almost decoupled hidden sector within a classically scale invariant theory\\}

\vspace{2 cm}

\author{Robert Foot}
  \email[Email: ]{rfoot@unimelb.edu.au}
   \affiliation{School of Physics, Research Centre for High Energy Physics,\\
    University of Melbourne, Victoria 3010, Australia.}
\author{Archil Kobakhidze}
  \email[Email: ]{archilk@unimelb.edu.au}
   \affiliation{School of Physics, Research Centre for High Energy Physics,\\
    University of Melbourne, Victoria 3010, Australia.}
\author{Kristian L. McDonald}
  \email[Email: ]{klmcd@triumf.ca}
   \affiliation{Theory Group, TRIUMF, 4004 Wesbrook Mall, Vancouver, BC V6T2A3, Canada.}
\author{Raymond R.Volkas}
\email[Email: ]{raymondv@unimelb.edu.au}
\affiliation{School of Physics, Research Centre for High Energy Physics,\\
    University of Melbourne, Victoria 3010, Australia.}
\smallskip

\begin{abstract}
\begin{center}
{\large Abstract}
\end{center}
If scale invariance is a classical symmetry then both the Planck scale and the weak scale should emerge as quantum effects. We show that this can be realized in simple scale invariant theories with a hidden sector.  The weak/Planck scale hierarchy emerges in the (technically natural) limit where the hidden sector decouples from the ordinary sector.  The weak scale is protected from quadratic divergences because of classical scale invariance, so there is no hierarchy problem. 
\end{abstract}
\maketitle

\bigskip

\bigskip

\baselineskip=16pt
\newpage

\section{Introduction}

One way to stabilize the electroweak scale against radiative corrections
is to promote scale invariance
to a good classical symmetry by eliminating the $\mu^2$ mass parameter
from the Higgs potential~\cite{Bardeen:1995kv}.
As first discussed by Coleman and E. Weinberg~\cite{Coleman:1973jx},
electroweak symmetry breaking can still arise
since scale invariance is anomalous and the radiatively corrected Higgs
potential may induce
spontaneous symmetry breaking. When applied to the minimal standard
model this elegant idea 
fails as the heavy top quark precludes the required dominance of
the bosonic contributions to the effective potential.

Additional bosonic contributions, in the form of extra gauge bosons
and/or scalars, can lead
to phenomenologically successful 
models~\cite{Hempfling:1996ht,Meissner:2006zh,Espinosa:2007qk,Chang:2007ki,Foot:2007cq,Hambye:2007vf}
of electroweak symmetry breaking - with
the electroweak scale stabilized against radiative corrections. A
pressing issue though, is what
to do about gravity, which of course, also involves a scale - the Planck
mass.
One possibility is that gravity can be treated separately from the other
interactions, and we need not
worry that the 
gravitational interaction explicitly violates scale invariance.
Alternatively, it is tempting to apply scale invariance to the the whole
fundamental theory -- including 
gravity. 

We propose that 
the Planck mass arises from the vacuum expectation value (VEV) of a
scalar field $S$~\cite{Zee:1978wi}, which is dynamically generated
via the radiative breaking of scale invariance.\footnote{Note that
demanding classical scale invariance motivates the absence of the usual
Einstein-Hilbert Lagrangian $\mathcal{L}\propto \sqrt{-g}M_{Pl}^2R$, which must
otherwise be arbitrarily omitted in models which spontaneously generate the Planck scale~\cite{Zee:1978wi}.} Thus we include
gravity by postulating 
the scale-invariant interaction term
\begin{eqnarray}
{\cal L} = \sqrt{-g}\left[ \frac{1}{2} \xi S^2 R \right].
\label{1}
\end{eqnarray}
Standard gravity emerges if $\xi \langle S \rangle^2 \equiv M_{Pl}^2$,
which requires $\langle S \rangle \sim M_{Pl}$ for $\xi \sim 1$ [here $M_{Pl} \equiv (8\pi G)^{-1/2} 
= 2.436 \times 10^{18}$ GeV is the reduced Planck Mass]. Due to
its large VEV,
$S$ should not couple directly to the standard particles, so we require
it to be an $SU(3)_c \otimes SU(2)_L \otimes U(1)_Y$
gauge singlet. 

Observe that the scalar $S$ only couples to the ordinary
particles through the 
Higgs quartic coupling,\footnote{There are also indirect couplings which
are due to kinetic mixing 
of $S' \equiv S-\langle S \rangle$ with the trace part of the graviton
field, as we discuss later.}
\begin{eqnarray}
\lambda_X \phi^{\dagger}\phi S^2,
\end{eqnarray}
where $\phi$ is the standard model Higgs doublet.
In the limit $\lambda_X \to 0$ the scalar potential separates:
\begin{eqnarray}
V(\phi, S) = V(\phi) + V(S).
\end{eqnarray}
In this limit, $V(\phi)$ is simply the Coleman-Weinberg potential, and 
given the heavy top quark, spontaneous symmetry breaking in $V(\phi)$
does not
arise. Thus, we end up with a massless Higgs particle. The $V(S)$ part
can undergo spontaneous symmetry breaking (depending on the particle
content in the `hidden sector'), leading to $\langle S \rangle \neq 0$,
and this VEV sets the Planck scale and gravity via Eq.(\ref{1}).
If $\lambda_X$ is small, but non-zero, then the symmetry breaking
can be communicated to the electroweak sector. In fact it is also
possible to have the $\lambda_X$ term 
induce symmetry breaking in {\it both} $\phi$ and $S$.
Either way,
the ratio of scales $\langle \phi \rangle/\langle S \rangle$ is
controlled by just the one adjustable parameter $\lambda_X$. 
Furthermore, the physically interesting limit,
where $\langle \phi \rangle/\langle S \rangle \sim M_W/M_{Pl} \to 0$,
corresponds to
the technically natural\footnote{In the absence of gravity, decoupling a sector increases 
the symmetry of the theory, because the action $S = \int d^4x {\cal L}_{\rm vis}(x) + 
\int d^4x' {\cal L}_{\rm hid}(x')$ is invariant under independent Poincar\'{e} transformations for the
visible and hidden sectors.  In the presence of gravity, this is no longer the case.  
However, as discussed further below, we are treating gravity classically in this analysis, so
radiative corrections due to quantum gravity are ignored.  We are assuming that these quantum gravity
effects are sufficiently small so as to not spoil the technical naturalness arising from 
switching off the non-gravitational couplings between the sectors.}
limit of $\lambda_X \to 0$.  This would give a
technically natural
solution to the hierarchy problem. 

The outline of this paper is as follows: In Sec.~\ref{higgs_analysis} we examine the
radiatively corrected Higgs potential
of the simplest models which have the weak and Planck scales arising from
the 
anomalous breaking of scale invariance.  
As will be discussed there, such models feature a pseudo-Goldstone boson
(PGB) which
is associated with the breaking of scale invariance. This PGB is the
main observable
new physics predicted by the model.
In Sec.~\ref{ints} we investigate the coupling of this resultant 
pseudo-Goldstone boson to ordinary matter and discuss the constraints
from
experiments. We comment on the cosmological constant, which vanishes
classically within 
our model, in Sec.~\ref{cc} and conclude in Sec.~\ref{conc}.

\section{Higgs potential analysis}
\label{higgs_analysis}

We first consider the simplest possibility of the addition of one real
scalar, $S$, which couples to the curvature as in Eq.~(\ref{1}).  Some of this
discussion is very similar to that given in Ref.~\cite{Foot:2007cq}.

The tree level scalar potential is
\begin{eqnarray}
V = {\lambda_{\phi} \over 2} (\phi^{\dagger} \phi)^2 
+ {\lambda_S \over 8} S^4 - {\lambda_X \over 2} (\phi^{\dagger}\phi)S^2,
\label{a1}
\end{eqnarray}
and we parameterize the fields in unitary gauge as
\begin{eqnarray}
\phi = {r \over \sqrt{2}} \left( \begin{array}{c}
0 \\
\sin \omega 
\end{array} \right),\quad \ S = r \cos \omega.
\label{a2}
\end{eqnarray}
In this parameterization, the potential of Eq.~(\ref{a1}) is rewritten as
\begin{equation}
V_0(r, \omega)=r^4\left(\frac{\lambda_{\phi}}{8}\sin^4\omega
+\frac{\lambda_S}{8}\cos^4\omega -
\frac{\lambda_X}{4}\sin^2\omega\cos^2\omega  \right).
\label{a3}
\end{equation}
The radial component $r$ of the Higgs fields, Eq.~(\ref{a2}),
factors out due to the absence of the tree-level mass parameter. This classical potential 
receives quantal radiative corrections in the manner of Coleman and 
Weinberg~\cite{Coleman:1973jx}.  We shall work in the parameter regime where the
1-loop-level correction is sufficiently accurate.
In general, the minimization of even a 1-loop-corrected
effective potential involving multiple scalars cannot be done
analytically, so we instead follow the approximation
scheme introduced in \cite{Gildener:1976ih} which is suitable in weakly
coupled scale-invariant theories.  

Following \cite{Gildener:1976ih}, we first ignore the perturbatively small
radiative corrections and minimize the tree-level potential of Eq.~(\ref{a3}). 
Taking the parameter regime $\lambda_X > 0$, letting $\langle r \rangle$ be
nonzero but arbitrary, the minimum is
\begin{eqnarray}
{\langle \phi \rangle^2 \over \langle S \rangle^2} &\equiv & {\langle
\tan^2\omega \rangle \over 2} =
\frac{1}{2} \frac{\lambda_X(\Lambda)}{\lambda_{\phi}(\Lambda)}~, \nonumber \\
\label{a6}
\end{eqnarray}
with
\begin{eqnarray}
\lambda_X (\Lambda) = \sqrt{ \lambda_{\phi} (\Lambda) \lambda_S
(\Lambda)}\ .
\label{r1}
\end{eqnarray}
The relation, Eq.~(\ref{r1}), is not a fine-tuning but rather the
definition of a renormalization point $\mu=\Lambda$ \cite{Gildener:1976ih},
where the running coupling constants depend on $\mu$ and $\Lambda$
is the specific value where Eq.~(\ref{r1}) holds.  Equation (\ref{r1}) trades a 
dimensionless parameter for the dimensionalful
renormalisation point $\Lambda$: the phenomenon of dimensional transmutation. 

The classical
potential has a flat direction described by the arbitrary $\langle r \rangle$
along the vacuum solution, which shall be removed by the radiative corrections.
Since the classical potential is zero along that direction, the one-loop correction 
necessarily dominates there.

We shall be interested in the limit where 
\begin{eqnarray}
\tan^2 \omega = {2\langle \phi \rangle^2 \over \langle S \rangle^2} \sim
{M_W^2 \over M_{Pl}^2} \to 0.
\end{eqnarray}
From Eqs.~(\ref{a6}) we see that this limiting case
occurs when $\lambda_X \to 0$. (The other
choice $\lambda_\phi \to \infty$ is not phenomenologically viable.)
Observe that $\lambda_X \to 0$ corresponds to
the limit where the hidden sector
completely decouples from the ordinary sector. As noted earlier, this is   
technically natural since all non-gravitational radiative corrections to
$\lambda_X$ must vanish in the limit where the hidden sector is completely decoupled. 

To calculate the tree-level masses, we expand the Higgs
potential, Eq.~(\ref{a1}),
around the vacuum: $\phi = \langle \phi \rangle + \phi'$, $S = \langle S
\rangle + S'$. 
Of the two physical scalars, only one gains mass at tree-level; the other is
classically massless due to the flat direction. Calling the state that
develops mass at tree-level $H$, 
and the tree-level-massless state $\sigma$ (the
pseudo-Goldstone Boson (PGB) of broken scale invariance), we obtain
\begin{eqnarray}
M_H^2 = \lambda_{\phi} v^2 + \lambda_X v^2 \ , \quad 
H = -\cos\omega \phi'_0 + \sin\omega S',
\label{ee}
\end{eqnarray}
(here $v\approx 246$ GeV is the electroweak VEV) and the PGB field is $\sigma = \sin\omega \phi'_0 + \cos\omega S'$. 
Note that in the $\lambda_X \to 0$ limit, the mass and interactions of
$H$ reduce
to the standard electroweak Higgs particle. (In the above we took $\lambda_X >0$.  If
we had made the opposite choice, unsuitable for our present purposes, then the usual
electroweak Higgs would have been the PGB \cite{Foot:2007cq}.)

To calculate the mass of the PGB boson $\sigma$, we add the one-loop 
correction to the tree-level
potential of Eq.~(\ref{a3}) along the ``radial'' flat-direction.  It has 
the form~\cite{Gildener:1976ih}\footnote{The calculation is performed within the framework of dimensional regularization 
of divergent integrals with the $\overline{MS}$ substraction scheme. The use of dimensional regularization is absolutely crucial in scale-invariant theories, since unlike other regularizations, it requires counterterms that preserve the form of the bare Lagrangian.  Emergence of  quadratic and  quartic divergences in other regularization schemes (e.g. cut-off or Pauli-Villars regularizations) must be regarded as artifacts of those regularization schemes because they require counterterms which are quadratic  and quartic in the regulator scale (i.e. the form of the counterterms is different form that of  the bare Lagrangian). }
\begin{equation}
\delta V_{\rm 1-loop}= Ar^4 \ + \
Br^4\log\left(\frac{r^2}{\Lambda^2}\right )~,
\label{14}
\end{equation}
where
\begin{equation}A=\frac{1}{64\pi^2 \langle r \rangle^4}\left[3{\rm
Tr}\left(M_V^4\ln\frac{M_V^2}{\langle
r\rangle^2}\right)+{\rm
Tr}\left(M_S^4\ln\frac{M_S^2}{\langle
r\rangle^2}\right)-4{\rm
Tr}\left(M_F^4\ln\frac{M_F^2}{\langle r\rangle^2}\right)
\right ]~,
\label{15}
\end{equation}
and 
\begin{equation}
B=\frac{1}{64\pi^2 \langle r \rangle^4}\left [3{\rm Tr}M_V^4+{\rm
Tr}M_S^4-4{\rm
Tr}M_F^4
\right ]~.
\label{16}
\end{equation}
The traces sum over all degrees of
freedom, with $M_{V,S,F}$ being the tree-level masses for
vectors, scalars and fermions, respectively.

The stationary condition $\frac{\partial \delta V_{\rm 1-loop}}{\partial
r}|_{r=\langle r \rangle}=0$  yields
\begin{equation}
\log\left(\frac{ \langle r \rangle}{\Lambda}
\right)=-\frac{1}{4}-\frac{A}{2B}.
\label{17}
\end{equation}
Computing the PGB mass, and using Eq.~(\ref{17}), one finds \cite{Gildener:1976ih}:
\begin{eqnarray}
M_{\sigma}^2  & = & \left. \frac{\partial^2 \delta V_{\rm 1-loop}}{\partial r^2
}
\right|_{r = \langle r \rangle} = 8 B \langle r \rangle^2  \nonumber \\
& = & 
\frac{1}{8\pi^2 \langle r \rangle^2}\left [3{\rm Tr}M_V^4+{\rm
Tr}M_S^4-4{\rm
Tr}M_F^4
\right ]~.
\label{18}
\end{eqnarray}
Evaluating the traces, we obtain
\begin{eqnarray}
M_{\sigma}^2 &\simeq & {1 \over 8\pi^2 \langle r \rangle^2} \left[M_H^4
+6M_W^4 + 3M_Z^4 - 12 M_t^4\right]\ . 
\label{a9}
\end{eqnarray}
The constraint $M_{\sigma}^2 > 0$, implies that
$M_H^4 >  12M_t^4 - 6M_W^4 - 3M_Z^4$,
evaluated at the scale $\Lambda \sim M_{W}$. Such a large
Higgs mass is
in conflict with the precision electroweak data. Thus additional scalar
or vector bosons are required
to make the model phenomenologically viable. However 
this is neither unexpected nor unwelcome since
the model as it stands does not explain neutrino masses and dark matter.

While there are many ways to complete the theory to achieve those goals,
it is not the purpose of the present paper to survey all the possibilities.
Instead, we shall be satisfied with an example.  We could, for instance, 
add a complex Higgs triplet, $\Delta \sim (1, 3, -2)$, to ensure 
phenomenologically-viable electroweak symmetry breaking 
and induce non-zero Majorana
neutrino masses~\cite{Foot:2007ay}. The most general
Higgs potential includes the terms
\begin{eqnarray}
{\cal L} = \lambda \Delta^{\dagger}\Delta S^2 + \lambda' \phi \Delta
\phi S + \mathrm{H.c.}\ .
\label{delta_terms}
\end{eqnarray}
The first term induces a mass for $\Delta$ whilst the second 
induces a VEV. This leads to neutrino masses via the coupling
\begin{eqnarray}
{\cal L} = \lambda_{\nu} \bar f_L \Delta (f_L)^c +\mathrm{H.c.}
\end{eqnarray}
where $f_L$ denotes a leptonic $SU(2)_L$ doublet and the superscript
denotes CP conjugation. If the radiative 
contributions from $\Delta$ dominate over the other radiative
contributions in the 
effective potential, one expects the PGB mass to be of order
\begin{eqnarray}
M_{\sigma} &\approx & {\sqrt{3\xi} \over 2\pi}{M_{\Delta}^2\over
M_{Pl}} \nonumber \\
 &\sim & \left({M_{\Delta} \over TeV}\right)^2 10^{-4}\ {\rm eV}.
\end{eqnarray}
Thus, we expect the mass of the PGB to range from around $10^{-4}$
eV to $\sim M_{Pl}$,
depending on the mass of $\Delta$. The hidden sector now decouples in
the limit that $\lambda_X$, $\lambda$ and 
$\lambda'$ vanish. Note that only the $\lambda'$ term in
(\ref{delta_terms}) breaks 
lepton number so the limit $\lambda'\rightarrow 0$, in which the
VEV of $\Delta$ also becomes suppressed, 
is technically natural. Provided $\lambda_X$ and $\lambda$ are both
small, the hidden sector 
remains weakly coupled and the weak/Planck hierarchy will be preserved. 
Thus one expects $\sigma$ to be light, and therefore observable, in the
decoupling limit.

Besides the above important phenomenological motivation for the extension of the minimal 
framework of our model, there is a purely theoretical one as well. The fact that the electroweak Higgs boson must be heavy ($M_H\gtrsim (12)^{1/4}M_{t}\approx 327$ GeV) and $\lambda_X$ is hierarchically small, implies 
[see Eq. (\ref{ee})] that the Higgs self-interaction coupling is strong $\lambda_{\phi}\gtrsim\sqrt{3}$ at the electroweak scale.  
This coupling becomes non-perturbative at energies below the Planck mass. While it is not immediately obvious whether the 
existence of this Landau pole for $\lambda_{\phi}$ spoils the weak/Planck scale hierarchy, it is certainly desirable to 
have a  perturbative theory at least at energies up to the Planck mass. In many extensions of the bosonic  sector of our 
minimal model the Landau pole can be pushed beyond the Planck scale, and, in some cases, completely removed.  For example, in the case of the electroweak triplet Higgs discussed above, with $M_{\Delta}>M_{H}$,  
the theory can be kept perturbative at all energy scales up to the Planck scale (and perhaps beyond) 
provided the electroweak Higgs boson is relatively light, $M_H\lesssim 120$ GeV or so.

A few clarifying remarks on the calculations in this section are in order. 
Obviously a completely consistent approach to the problem requires the
quantization of gravity. However, since classical general relativity (or
its scalar-tensor extensions) does not readily admit a sensible quantum
theory, we have taken in the above the widely accepted point of view
that the metric may be treated as a classical background field.
Moreover, we have taken this background to be (nearly) flat,
$g_{\mu\nu}(x)\approx\eta_{\mu\nu}$, and performed calculations in flat
spacetime. This is justified if the inverse of the local curvature
radius $a$ is much smaller than the typical mass scale of the problem,
because only infrared modes ($k < 1/a$) of fields are influenced by the
curvature, while the quantum effects are dominated by the ultraviolet
modes ($k \gg 1/a$). Under these assumptions the above analysis is
justified.

Let us summarize the main point concerning the hierarchy problem~\cite{Bardeen:1995kv} (see also~\cite{Meissner:2007xv}). The classical scale invariance does remove all divergences that go as  powers of the regulator mass scale. Consequently, no large (quadratic) corrections to the electroweak mass scale are expected and this explains its smallness  in a technically-natural way (we have no fundamental explanation for the smallness of certain couplings  though). In this regard, scale invariance plays the same role in the solution of the hierarchy problem as the more popular softly broken supersymmetry. Extensions of the scalar sector,  which are anyway necessary to incorporate a fully realistic phenomenology (e.g. neutrino masses, dark matter, etc.), as briefly discussed above, cannot reintroduce the hierarchy problem.  
In these extensions,  Landau poles associated with some asymptotically non-free coupling constants  can be pushed beyond the Planck scale, so that the standard quantum field theory, in the domain of its applicability,  is in the perturbative regime. Thus, if some fundamental theory, incorporating also a quantum theory of gravitation, possesses scale invariance in its low-energy domain (where the quantum gravitational effects decouple), the hierarchy problem can be eliminated in the way we have discussed in this section.

\section{Interactions of the PGB with ordinary matter}
\label{ints}

The new scalar $\sigma$ which couples non-trivially with the curvature
$R$ is the main new physics
predicted in these models. This new physics will manifest itself
though modifications to gravity at scales
\begin{eqnarray}
d \lesssim {1 \over M_{\sigma}}\ .
\end{eqnarray}
If $\sigma$ is light, $M_{\sigma}\stackrel{<}{\sim}$ eV, then this
effect can be experimentally probed in tests 
of Newton's inverse square law at short distance scales.
To determine the magnitude of this effect, we need to compute
the coupling of the PGB scalar $\sigma$ 
with ordinary matter.

Consider the field $S'$, representing the fluctuations around the vacuum.  In the weak
gravity limit, $R \to 0$, the gravitational field equations can be
linearized, leading to~\cite{review}
\begin{eqnarray}
{M_{Pl}^2 \over 2} [ - \Box h_{\mu \nu} + ...] +
\frac{2M_{Pl}^2}{\langle S \rangle}
(\eta_{\mu \nu} \Box - \triangledown_{\mu}\triangledown_{\nu}) S' = 
T^m_{\mu \nu}\ ,
\end{eqnarray}
where $T^m_{\mu \nu}$ is the stress-energy due to matter.
Thus $S'$ and the trace of $h_{\mu \nu}$ are kinetically mixed. The
kinetic term can be
diagonalized by introducing the field
\begin{eqnarray}
\tilde{h}_{\mu\nu} \equiv h_{\mu \nu} - \frac{1}{2} \eta_{\mu \nu} h -
2\eta_{\mu\nu} \frac{S'}{\langle S \rangle}\ .
\end{eqnarray}
This transformation to the kinetically diagonal basis,
($\tilde{h}, S')$, 
introduces a host of exotic Planck scale derivative couplings.
In the Higgs potential analysis of the preceding section we have ignored
this kinetic mixing, essentially
doing perturbation theory in the
kinetically mixed basis ($h_{\mu \nu}, S'$). 
Since we are treating gravity as a classical background field, such a
procedure is reasonable.
However, to determine the couplings of $\sigma = \sin\omega \phi'_0 +
\cos\omega S'$ to matter it is necessary to
transform to the kinetically diagonal basis.
In the kinetically diagonalized basis ($\tilde{h}, S'$),
 the kinetic term for for the $S'$ field has the form
 \begin{equation}
 \frac{1}{2}(1+6\xi)\partial_{\mu}S' \partial_{\mu}S'~.
 \label{c}
 \end{equation}
 
 We are primarily interested in the leading order interactions of the
PGB $\sigma$ with light matter fermions, 
 which are linear in $\sigma$. Since the kinetic terms of the fermions
(as well as vector bosons) are conformally 
 invariant, these interactions arise at the classical level through the
fermion mass terms, 
 \begin{eqnarray}
 \sqrt{-g}m_f\bar\psi_f\psi_f \to 
 \left(1+\frac{1}{2}h-\frac{4\sigma}{\langle r\rangle}\right)
 \left(1+\frac{3 \sigma}{\langle r\rangle}\right )m_f\bar\psi_f \psi_f =
\nonumber \\
 m_f \bar \psi_f \psi_f - \frac{m_f}{\langle r\rangle}\sigma
\bar\psi_f\psi_f~.
 \label{d}
 \end{eqnarray} 
However, the $\sigma$ interaction term is exactly cancelled by the
Yukawa interactions induced by the 
Higgs $(\phi)-S$ mass mixing, 
$\frac{y_f}{\sqrt{2}}\sigma \sin\omega \bar\psi_f\psi_f =
+\frac{m_f}{\langle r\rangle}\sigma \bar\psi_f\psi_f$.  
This remarkable cancellation is a manifestation of the underlying scale
invariance and 
ensures that fundamental fermions are only weakly coupled to the PGB. 

The non-trivial admixture of gluons inside nucleons requires us to
consider $\sigma$-gluon interactions induced at the 
quantum level in order to examine the PGB coupling to matter. As the
Weyl rescaling is anomalous we must include 
the anomalous trace of the energy-momentum tensor in the effective
Lagrangian. Thus
\begin{equation}
\sqrt{-g}T^{\rm anom}\to -4\frac{\sigma}{\langle r \rangle}\left(
\frac{\beta(g_i)}{g_i}F^{a}_{\mu\nu}F^{a~\mu\nu}+\gamma(m_f)\bar
\psi_f\psi_f+... \right)~,
\label{e}
\end{equation}
where $\beta(g_i)$ and $\gamma(m_f)$ are respectively the
$\beta$-functions and the anomalous dimensions (here $i=1,2,3$ correspond to the $SU(3), SU(2), U(1)$ coupling constants,
with $F^{a}_{\mu \nu}$ the corresponding field strength tensor). 
The dominant interaction in (\ref{e}) results from the gluon fields,
from which we can deduce the effective 
interaction of the PGB with nucleons~\cite{shifty}, 
\begin{eqnarray}
{\cal L}_{eff} \approx
-\frac{4}{\sqrt{1+6\xi}}\frac{m_N}{\langle r \rangle}\sigma \bar N N~,
\label{f}
\end{eqnarray}
where we have used the result~\cite{shifty} that
\begin{eqnarray}
\langle N|\frac{\beta(g_s)}{g_s}F^{a}_{\mu\nu}F^{a~\mu\nu}|N\rangle \approx m_N \bar \psi_N \psi_N~,
\label{g}
\end{eqnarray}
and we have assumed canonical normalization for the $\sigma$ field
(see Eq.~(\ref{c})). Thus the interchange of a $\sigma$ field between
two nucleons gives rise to a static potential between two
(non-relativistic) nucleons, 
\begin{equation}
V(r)=G_{N}\frac{32\xi}{1+6\xi}m_{N}^2\frac{{\rm e}^{-M_{\sigma}r}}{r}
\label{h}
\end{equation}
where $G_{N}$ is Newton's constant. This has to be compared with
the experimentally measured gravitational potential between two nucleons. 
Recent torsion-balance experiments \cite{Kapner:2006si} showed no
deviation from the standard gravitational 
inverse-square law, and this implies the bound,  
\begin{equation}
\xi \lesssim 3\times 10^{-4} \ {\rm for} \ M_{\sigma} \sim 10^{-4} \ {\rm eV}\ .
\end{equation}
However, for $M_{\sigma} \stackrel{>}{\sim} 10^{-2}$ eV (corresponding to a
distance 
scale of $\sim 10^{-3}$ cm), the limit on $\xi$ is completely
relaxed.
Importantly the current experiments are probing the 
interesting parameter range $M_{\sigma} \sim 10^{-2}-10^{-4}$ eV, $\xi
\sim 1$, and thus forthcoming experiments may help in testing this
theory.

\section{Cosmological constant}
\label{cc}

Promoting scale invariance to a good classical symmetry requires the
cosmological  
constant to vanish classically. In other words in this type of
scale-invariant theory, we would expect
the cosmological constant to arise only as a radiative effect. 
That the cosmological constant vanishes at tree-level is an interesting
feature of these
theories, which optimistically speaking, might ultimately lead to an
understanding of its currently small value
inferred from the accelerated expansion rate of the Universe. 

The lowest order contribution to the cosmological constant from the
scalar sector is negative and 
given by~\cite{Gildener:1976ih}:
\begin{equation}
E_{Higgs} = -\frac{1}{2} B \langle r \rangle^4.
\label{higgs2}
\end{equation}
Observe that this scalar vacuum energy is proportional to the one-loop
contribution to the PGB mass [Eq.~(\ref{18})] and is uniquely determined
once the particle content of the model is specified.

An additional independent contribution to the vacuum energy-density results from the gluon condensate,
which is also negative \cite{vain}: 
\begin{eqnarray}
E_{QCD} = - (9/32) \langle 0 | \frac{\alpha_s}{\pi}F_{\mu \nu}^a F^{a~\mu \nu} |0\rangle 
\end{eqnarray}
Numerically, $\langle 0| (\alpha_s/\pi)F_{\mu \nu}^a F^{a~ \mu \nu} |0\rangle  \simeq 0.012 \ {\rm GeV^4}$.

Note that since the contribution to the vacuum energy from the effective potential and QCD vacuum
are the same sign one cannot arrange a fine-tuned cancellation 
between them to produce the observed value. One can imagine a scenario where some additional source of the 
cosmological constant arises to cancel these terms (perhaps from the hidden sector). A particularly 
interesting case arises if this exotic source is of the same order as the gluon condensate (MeV). 
Then the contribution from the scalar sector should vanish to order MeV and 
Eq. (\ref{16}) implies an approximate mass relation between the new fields which dominate the one loop 
potential and the top quark.

\section{Conclusion}
\label{conc}

We have considered the possibility that scale invariance is a classical
symmetry of all fundamental interactions, including 
gravity. Scale invariance is broken radiatively which generates standard
electroweak symmetry breaking as well
as the Planck scale via the VEV of a hidden sector scalar. 
The hierarchy problem in
such theories is  eliminated, with the weak/Planck scale 
hierarchy emerging in the (technically natural) limit where the hidden
sector
decouples from the ordinary sector.

\vskip 1cm
\noindent
 {\bf Acknowledgments:} This work was supported by the Australian
Research Council.

\end{document}